\documentclass[preprint2]{aastex}
\usepackage{emulateapj5}
\usepackage{onecolfloat5}

\newcommand{\kms}{km~s$^{-1}$}
\newcommand{\water}{H$_2$O~}
\newcommand{\jyb}{Jy~beam$^{-1}$}

\slugcomment{Accepted for publication in ApJ}

\shorttitle{Magnetic fields in evolved stars}
\shortauthors{Vlemmings et al.}

\begin{document}

\twocolumn[
\title{Magnetic Fields in Evolved Stars: Imaging the Polarized Emission of High-Frequency SiO Masers}


\author{W. H. T. Vlemmings\altaffilmark{1}, E. M. L. Humphreys\altaffilmark{2} \& R. Franco-Hern{\'a}ndez\altaffilmark{1}}

\begin{abstract}
  We present Submillimeter Array observations of high frequency SiO
  masers around the supergiant VX~Sgr and the semi-regular variable
  star W~Hya. The $J=5-4, ~v=1~^{28}$SiO and $v=0~^{29}$SiO masers of
  VX~Sgr are shown to be highly linearly polarized with a
    polarization from $\sim5-60\%$. Assuming the continuum emission peaks at the stellar position, the masers are found within $\sim
  60$~mas of the star, corresponding to $\sim100$~AU at a distance of
  1.57~kpc. The linear polarization vectors are consistent with a large
  scale magnetic field, with position and inclination angles similar
  to that of the dipole magnetic field inferred in the H$_2$O and OH
  maser regions at much larger distances from the star. We thus show
  for the first time that the magnetic field structure in a
  circumstellar envelope can remain stable from a few stellar radii
  out to $\sim1400$~AU. This provides further evidence supporting the
  existence of large scale and dynamically important magnetic fields
  around evolved stars. Due to a lack of parallactic angle coverage,
  the linear polarization of masers around W~Hya could not be
  determined.  For both stars we observed the $^{28}$SiO and
  $^{29}$SiO isotopologues and find that they have a markedly
  different distribution and that they appear to avoid each
  other. Additionally, emission from the SO $5_5-4_4$ line was imaged
  for both sources. Around W~Hya we find a clear offset between the
  red- and blue-shifted SO emission. This indicates that W~Hya is
  likely host to a slow bipolar outflow or a rotating disk-like
  structure.
\end{abstract}
\keywords{Masers, Polarization, Stars: Individual (VX~Sgr, W~Hya), Stars: Late-Type}
]

\altaffiltext{1}{Argelander-Institut f{\"u}r Astronomie, University of Bonn, Auf dem H{\"u}gel 71, D-53121 Bonn, Germany}
\altaffiltext{2}{ESO, Karl-Schwarzschild-Str. 2, D-85748 Garching, Germany}

\section{Introduction}

\begin{figure*}
\epsscale{2.0}
\plottwo{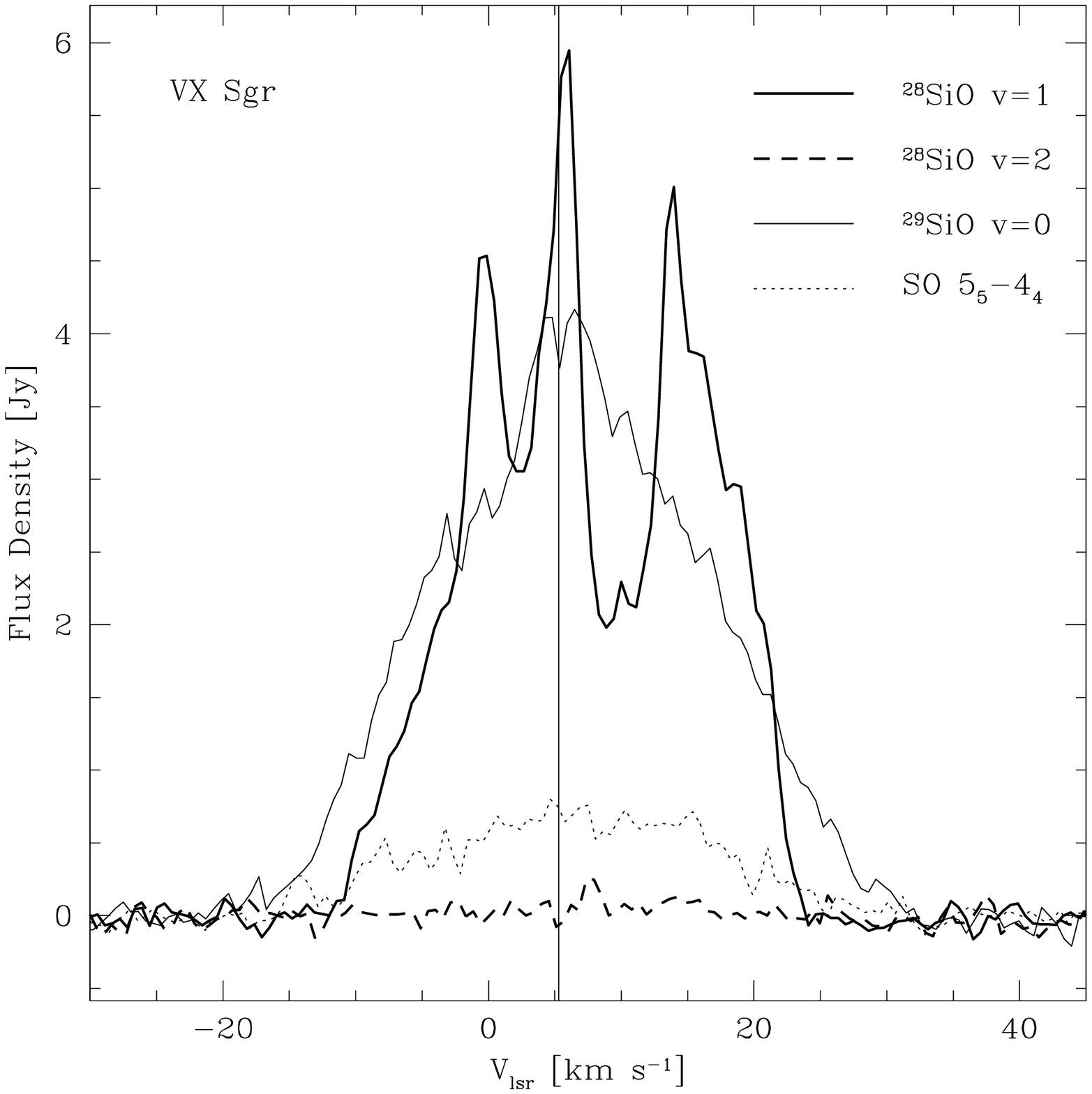}{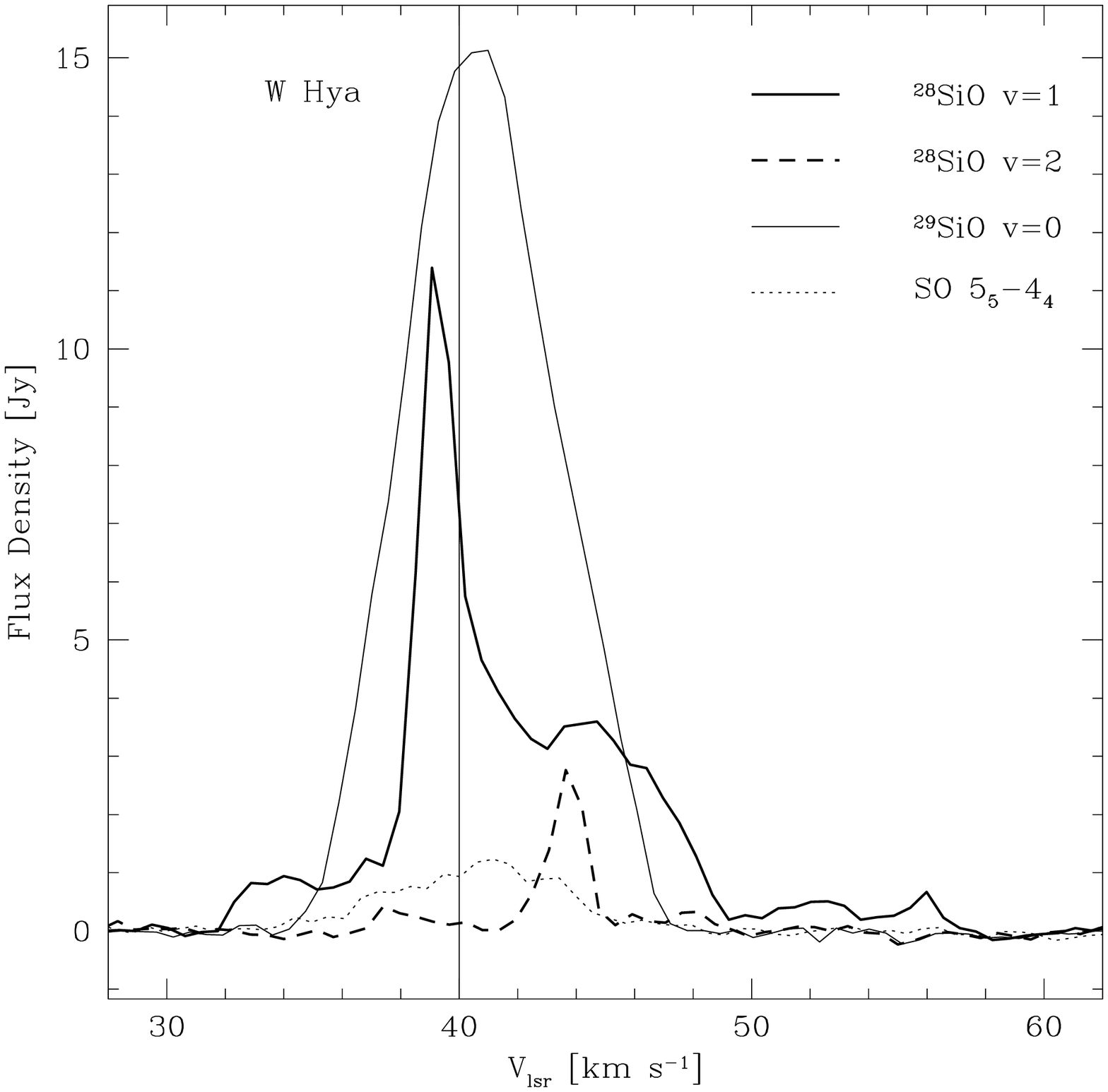}
\caption{SMA spectra of VX Sgr (left) and W Hya (right) for the $^{28}$SiO $v=1, J=5-4$ masers at 215.6~GHz (thick solid line), the $^{28}$SiO $v=2, J=5-4$ masers at 214.1~GHz (thick dashed line), the $^{29}$SiO $v=0, J=5-4$ masers at 214.4~GHz (thin solid line) as well as the SO $5_5-4_4$ line at 215.2 GHz (thin dotted line). The vertical lines denote the stellar velocity.} \label{spectra}
\end{figure*}

Many different SiO maser transitions have been observed around evolved
stars of different classes \citep[][and references
therein]{Kemball07}. High angular resolution observations of 43
and 86~GHz masers have shown that they occur within a few stellar radii
from the photosphere, in the dynamical region between the pulsating
star and the dust formation zone \citep[e.g.][]{Cotton04, Greenhill95,
  Chen06}. SiO masers are therefore excellent probes of the processes that
drive stellar mass loss, and that define the outflows that give
rise to potential asymmetries in the circumstellar envelope (CSE).

SiO maser emission from evolved stars has been detected from the $v=0$
to $v=4$, $J=1-0$ up to at least the $J=8-7$ transitions (e.g., Gray,
Humphreys \& Yates 1999; Pardo et al. 1998; Humphreys et al. 1997;
Jewell et al. 1987). The conditions giving rise to emission are
typically n(H$_{2}$)=10$^{10\pm1}$ cm$^{-3}$ and T$_{k }$ $\gtrsim$
1500~K, with the dominant pumping mechanism, radiative or
  collisional, a matter of debate \citep[e.g.][]{Herpin00,
  Lockett92}. However, it is agreed that line overlaps between the SiO
main isotopologues ($^{28}$SiO, $^{29}$SiO and $^{30}$SiO) likely play
a role in the pumping of e.g., the very highly excited $v=4$ $J=5-4$
maser line ($\sim7000$ K above ground state;
\citealt{Cernicharo93}). SiO maser $v=1$ and $2$, $J=5-4$ emission is
commonly found toward Mira variables for which the lower-$J$ masers
have already been detected. High-frequency SiO maser emission in Miras
appears to be strongly tied to stellar pulsation, with the velocity of
the emission a function of phase, and this phase dependence correlated
with photon luminosity. Indeed, the $J=7-6$ and $8-7$ lines are weak
or absent from $\phi=0.4$ to $0.7$, unlike the low-$J$ masers
\citep{Gray99}. \citet{Humphreys99} finds that highly rotationally
excited SiO maser lines arise from a subset of the physical conditions
leading to $J=1-0$ emission, and are strongest in dense, warm
post-shock gas.

Besides probing the dynamics and physical conditions in the CSEs,
masers are also good probes of the magnetic field
\citep[e.g.][]{Vlemmings07}.  Most of the information on the magnetic field
around evolved stars comes from maser polarization
observations. Ordered magnetic fields with a strength of order a
mG have been detected in the OH maser regions at large distances from
the star \citep[e.g.][]{Szymczak98}. Closer in, H$_2$O masers
also indicate the presence of a dynamically important magnetic
field, with typical values of a few hundred mG
\citep[e.g.][]{Vlemmings05}. Finally, 43 and 86~GHz SiO maser observations 
reveal field strengths of several Gauss at only a few stellar radii from the star,
assuming a standard Zeeman interpretation
\citep{Kemball97, Herpin06}. The higher frequency SiO masers display a
large fractional linear polarization \citep[][hereafter S04]{Shinnaga04}, and this polarization is potentially a good probe of the
circumstellar magnetic field morphology. However, as these masers
likely exist in the regime where the Zeeman splitting is comparable to
the rate of stimulated maser emission, the interpretation of their
polarization is not straightforward \citep{Nedoluha94}. 

Here we present the polarization and distribution of the $v=1$ and $2$ $J=5-4$
$^{28}$SiO and $v=0, J=5-4$ $^{29}$SiO masers around the supergiant
VX~Sgr and the semi-regular variable star W~Hya observed with the
Submillimeter Array (SMA; \citealt{Ho04}). In the remainder of the paper we
will forgo most mentions of the rotational quantum numbers $J=5-4$
and only give the vibrational quantum number $v=0,1$ or $2$. In
addition to the maser observations, we present observations of the SO
$5_5-4_4$ line that was detected for both stars.

VX~Sgr is a red supergiant star, located at a distance of
$1.57\pm0.27$~kpc \citep{Chen07}. It has an optical variability period of
732~days which suggests, assuming the Mira period-luminosity relation, it
has a mass of $\sim10$~M$_\odot$. Its envelope hosts a number of
maser species, from which the stellar velocity is estimated to be
$V_{\rm lsr}=5.3$~\kms~\citep{Chapman86}.

W~Hya, at a distance of $98$~pc \citep{Vlemmings03}, is typically
classified as a semi-regular variable star as its period is observed to
vary between $\sim350-400$ days. However, it is also often
considered a Mira variable due to the regular shape and the
large amplitude of its light curve \citep[e.g.][]{Gomez86}. The
envelope of W~Hya harbors many different maser species, and its
OH maser spectrum indicates a stellar velocity of $V_{\rm
  lsr}=40$~\kms~\citep{Szymczak98}.

\begin{figure}
\epsscale{1.0}
\plotone{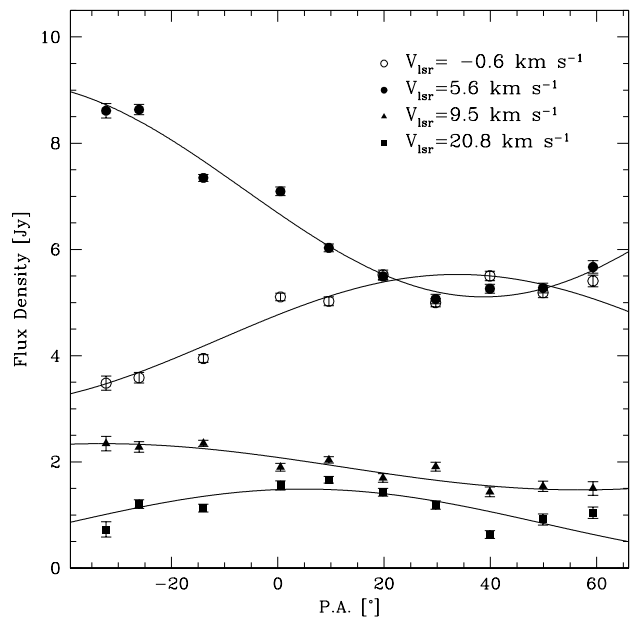}
\caption{The feed position angle (P.A.) vs. the apparent flux density for four of the $^{28}$SiO $v=1, J=5-4$ masers around VX~Sgr. The error bars denote the uncertainties on the flux density as derived by the {\it jmfit} Gaussian fitting procedure and do not contain potential systematic effects due to time variable errors in the gain calibration.} \label{vxpa}
\end{figure}

\section{Observations}
\label{obs}

The SiO masers of VX~Sgr and W~Hya were observed on UTC 2008 July 20
with the SMA in extended configuration. The resulting beam size was
$\sim1.5\times1.0$~arcseconds. The observations were centered at
214.8~GHz ($\lambda=1.4$~mm) to allow simultaneous observation of both
the $v=1$ and $v=2$ $^{28}$SiO maser lines at 215.5959 and
214.0885~GHz respectively in one 2~GHz band. This additionally covered
the frequencies of the $^{29}$SiO $v=0$ maser and the SO $5_5-4_4$
line. The frequency resolution was $400$~kHz, corresponding to
$\sim0.56$~\kms. The 2~GHz bandwidth, excluding the channels with line emission, was used to observe the
  dust continuum.  For the relative amplitude and phase calibration
of VX~Sgr we used NRAO530 ($\sim12.4^\circ$ away; $2.45$~Jy) and for
W~Hya we used 1337-129 ($\sim15.8^\circ$ away; $2.4$~Jy). For the
absolute amplitude calibration we used 3C454.3 with a flux density of
$22.63$~Jy. In both cases we used the brightest $v=1, ~^{28}$SiO maser
to self-calibrate the data. We estimate the absolute positions to be
accurate to $\sim0.2$~arcseconds, while the relative positional
accuracy of the different maser species and the dust should be good to
better than $\sim15$~mas. The data were initially calibrated using the
MIR IDL package, continuum subtraction and averaging was performed in
MIRIAD, and the imaging and phase self-calibration was performed in
AIPS. The channel rms noise is $\sim0.05$~\jyb.

After imaging and self-calibration we used the AIPS image-plane
  component fitting task {\it jmfit} to determine the position of the
maser emission in each individual velocity channel. When emission is
detected at $>10\sigma$, we define the fitted position in each channel
as a maser spot following the definition of \citet{Chen06}. A maser
feature is then a group of maser spots that occurs within a small
spatial and spectral region. Very Long Baseline Interferometry (VLBI)
SiO maser observations at high frequency resolution indicate that
there are typically several tens of such features with line-widths of
order $1$~\kms and a typical size of
$\sim1$~AU~\citep[e.g.][]{Chen08}. Both the spectral as well as the
spatial resolution of our observations are insufficient to identify
individual maser features. The maser spots defined here thus denote
the flux weighted average position of features that contribute flux in
the individual velocity channels. As the beam size (half-power
  beam width, HPBW) of our observations is
  $\sim1.5\times1.0$~arcseconds, the formal position errors on the
  masers depend on the signal-to-noise ratio (SNR) at which the maser
  spots were detected through $\Delta(\delta,\alpha)={\rm
    HPBW}/(2\times{\rm SNR})$.

As was performed by S04, we measured the linear polarization by
using earth rotation polarimetry, the diurnal rotation of the sky over
the SMA. As the SMA antennas have alt-azimuth mounts and fixed
linearly-polarized feeds at the Nasmyth focus, the position angle (P.A.) of the feeds on
the sky rotates during the observations according to Equation 1 of
S04:
\begin{equation}
{\rm P.A.}[^\circ] = 45^\circ - a - \sin \frac{\cos(\phi_{\rm lat})\sin(h)}{\cos(a)}.
\end{equation}
Here $a$ and $h$ are the elevation and hour angle of the source
respectively, and $\phi_{\rm lat}$ is the latitude of the SMA
($\sim19^\circ49'27"$). For VX~Sgr, $h$ ranged from $-2.3$ to $2.8$~hr
and the P.A. varied from $-35^\circ$ to $60^\circ$. For W~Hya, a
delayed start of the observations resulted in an hour angle range of
$0.6$ to $3.4$~hr and thus a smaller P.A. coverage from $15^\circ$ to
$75^\circ$.

\section{Analysis and Results}

\subsection{SiO Maser Spectra}

The spectra for the SiO maser transitions as well as the SO line are
shown in Fig.~\ref{spectra}. Around VX~Sgr, no significant emission was
detected for the $^{28}$SiO $v=2$ transition. The $v=1$ spectral
profile has significantly changed compared with the detection in 1986
by \citet{Jewell87} and the peak flux density decreased from $\sim15$ to
$\sim6$~Jy. The $^{29}$SiO emission is weaker, with a peak
flux density of $\sim4$~Jy and extends to slightly broader velocities with
respect to the star. It also shows fewer individual features in the
spectrum. Both lines peak close to the $V_{\rm lsr}=5.3$~\kms~ stellar
velocity.

The SiO maser emission of W~Hya is stronger than that of VX~Sgr in all
the three lines observed. Although the peak flux density has decreased by a factor
of 5, the $^{28}$SiO $v=1$ spectrum is remarkably similar to
that observed by \citet{Jewell87} over two decades earlier. It does
not, however, resemble the spectrum obtained in 1995
\citep{Humphreys97}. The shape of the $^{28}$SiO $v=2$ spectrum is
significantly different from that of the $v=1$ transition. Where the
$v=1$ line peaks around $V_{\rm lsr}=39$~\kms~ with a flux density of
$\sim11$~Jy, the $v=2$ maser is weaker ($\sim3$~Jy) and peaks at
$\sim44$~\kms. As was the case for VX~Sgr, the $^{29}$SiO maser spectrum has
less structure than that of the $^{28}$SiO masers, while it peaks
close to the stellar velocity of $V_{\rm lsr}=40$~\kms. At
$15$~Jy, it is stronger than the $^{28}$SiO lines.

\begin{figure*}
\epsscale{2.0}
\plottwo{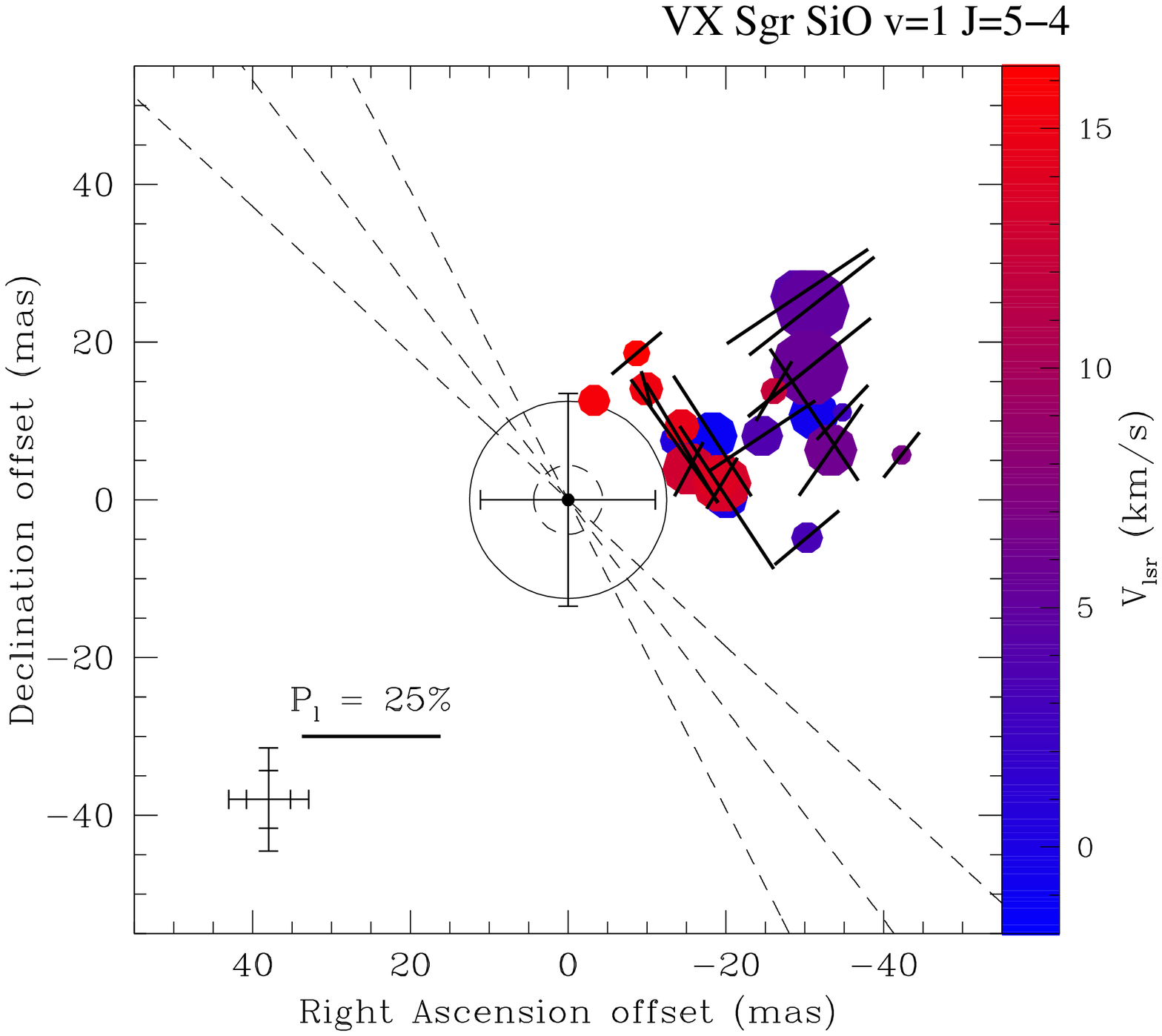}{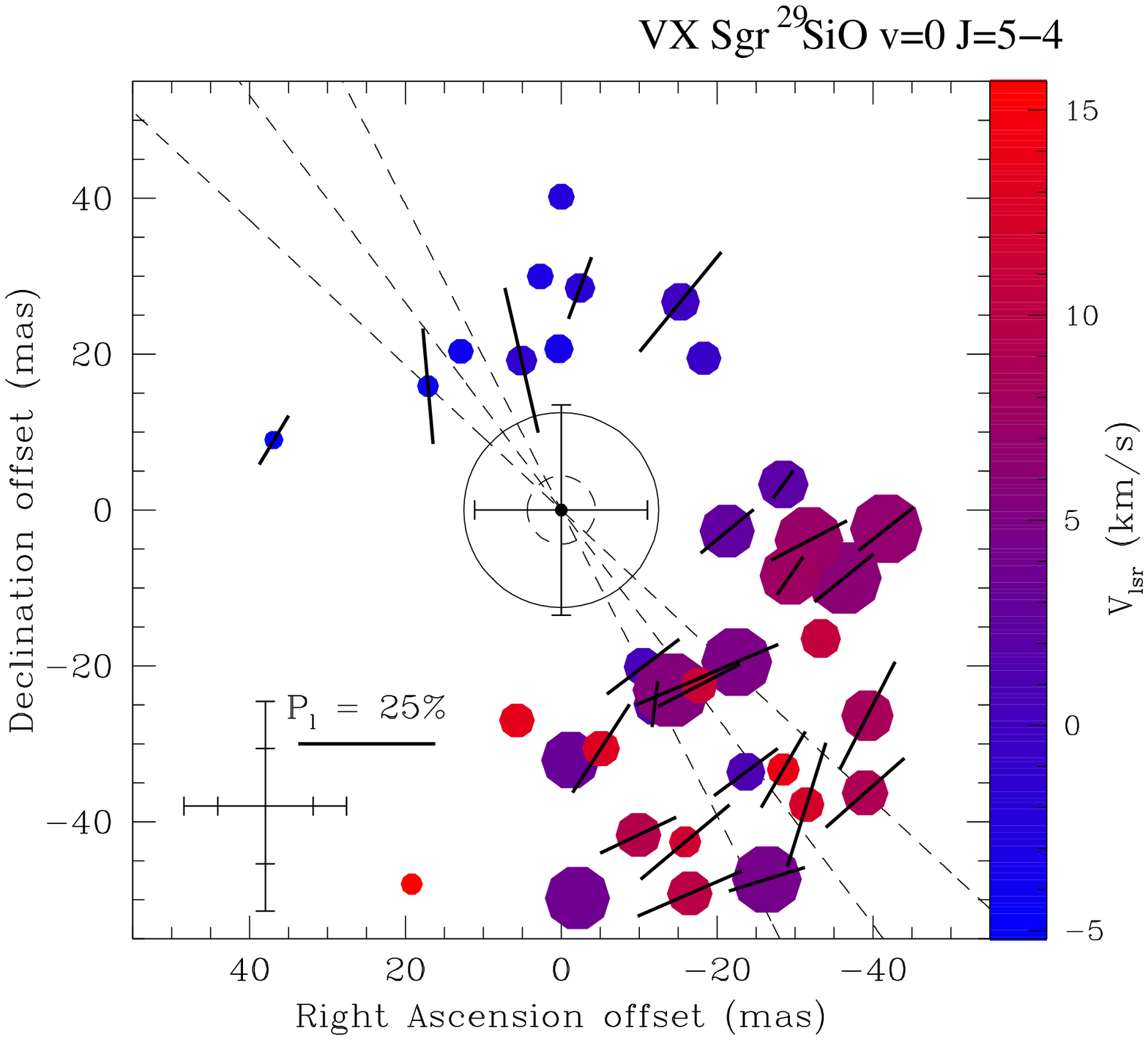}
\caption{Positions and polarization of the $v=1, J=5-4$ $^{28}$SiO (left) and $v=0, J=5-4$ $^{29}$SiO masers (right) around VX~Sgr. The masers spots are plotted with respect to peak of the dust emission ($\alpha_{\rm J2000}=18^h08^m04^s.068\pm0.014$ and $\delta_{\rm J2000}=-22^\circ13'28".416\pm0.200$).
The uncertainty in the dust position relative to the masers is indicated by the error bars in the center. Also note the difference in velocity scale between the two transitions. The symbols are linearly scaled according to brightness ($^{28}$SiO: $3.8$--$6.9$~\jyb; $^{29}$SiO: $2.3$--$4.2$~\jyb). For $^{28}$SiO, we only present the masers that were detected with a $SNR>50$. For the weaker $^{29}$SiO we plot the masers detected at $SNR>30$. The minimum and maximum positional errors for the plotted maser spots are indicated by the error bars in the bottom left. The black vectors are the observed polarization vectors scaled linearly according to polarization fraction. The long dashed inner circle indicates the star with a radius of $\sim4.35$~mas \citep{Monnier04} and the solid circle indicates the location of the 43~GHz SiO masers \citep{Chen06}. The dashed lines indicate the position angle and its uncertainty of the inferred orientation of the dipole magnetic field of VX~Sgr observed using H$_2$O and OH masers \citep{Vlemmings05, Szymczak01}.} \label{vxsio}
\end{figure*}

\subsection{Linear Polarization Analysis}

Because of the rotation of the polarized feeds on the sky as described in \S~\ref{obs}, the brightness $S$ of a linearly polarized source changes as a function of feed position angle according to the equation
\begin{equation}
S = S_0 + S_p \cos[2({\rm P.A.} - \chi)].
\end{equation}
Here $S_0$ is the average and $S_p$ the polarized brightness, and $\chi$ the electric-vector position angle (EVPA). The fractional linear polarization $P_l=S_p/S_0$. An example for four of VX~Sgr maser spots is shown in Fig.\ref{vxpa}. 

For a sufficient range of P.A. it is thus possible to determine the
EVPA and fractional polarization.  We performed a least
squares fit to all channels with significant ($>10\sigma$) maser
emission, having split the data in single 20~min blocks. The flux
density and associated uncertainties of the individual observing
blocks were determined using the AIPS task {\it jmfit}. Assuming a
constant dust continuum flux density during the observations, we determined an
initial gain factor for each of the time steps. The gain corrections
derived from the dust were $\lesssim6\%$, indicating that the dust was
not polarized at a level higher than $\sim6\%$. This, as well as the
non-detection of polarization in the SO line, indicates that the
observed SiO maser polarization is not an artifact of our
self-calibration and further reduction process. Still, as using the
dust for gain calibration could in principle introduce a systematic
bias due to possible dust polarization at a level of a few percent, we
also adopted the method described by S04 in an iterative minimization
routine. This yielded final gain corrections for the individual time
intervals of $2-9\%$. Still, remaining time variable uncertainties in
the flux calibration will due to, for example, variable phase
coherence add non-Gaussian errors to the flux density
determinations. The influence of remaining systematic errors on the
polarization fitting can be estimated using the bootstrap technique
\citep[e.g.][]{Efron91}. This technique allows determination of the
the fitted values and their associated uncertainties without relying
on Gaussian statistics and has previously been applied to assess
  image fidelity in radio interferometric polarimetry
  \citep{Kemball05, Kemball10}. The bootstrap technique consists of
the following; we treat the $N$ data points ${S_i}$ for each
  20-min observing interval as a set of independent estimates for the
flux density. We then generate data samples of size $N$ generated from
the original set with replacement, discarding samples with fewer than
four independent measurements. We generate 100,000 of such data
samples for which the least-squares fitting routine will determine
their best fitting $S_0$, $S_p$ and $\chi$. From the distribution of
these sample we then determine the most probable value for each of the
parameters as well as their most compact $68\%$ confidence interval.

\subsubsection{VX~Sgr}

\begin{figure}
\epsscale{1.0}
\plotone{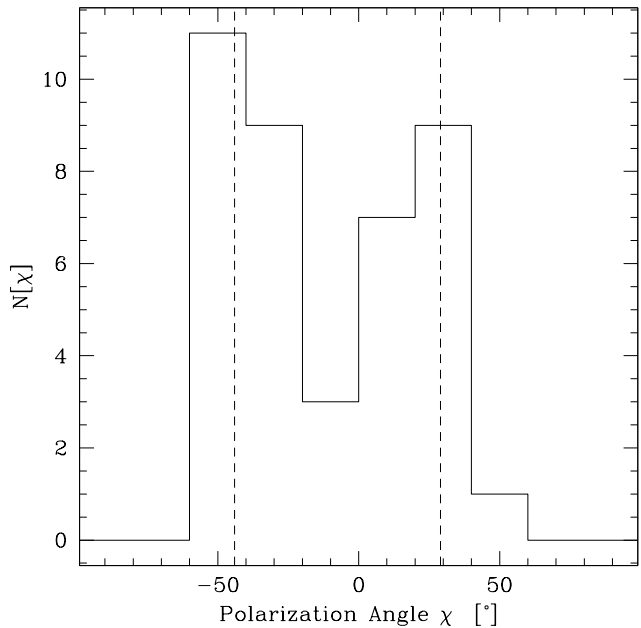}
\caption{The distribution of the electric-vector position angles ($\chi$) for the $v=1, J=5-4$ $^{28}$SiO masers of VX~Sgr. The dashed lines indicate the error weighted average peaks of the bimodal distribution of $\chi$. The histogram was constructed using all detected maser spots indicated in Table.1. } \label{vxdist}
\end{figure}

The polarization fractions and angles for the SiO masers of VX~Sgr are
given in the online Table.1 and shown in Fig.~\ref{vxsio}. The uncertainties are derived using the aforementioned bootstrap method. We find that
the average polarization fraction of the $^{28}$SiO $v=1$
masers $\langle P_l\rangle=26\pm16\%$ with the highest value reaching
almost $80\%$. The maser EVPA $\chi$ has, as seen in
Fig~\ref{vxdist}, a bimodal distribution, with peaks at
$\langle\chi\rangle=-44\pm13^\circ$ and
$\langle\chi\rangle=29\pm11^\circ$. The $^{29}$SiO masers have a
slightly lower average polarization $\langle
^{29}P_l\rangle=22\pm12\%$ with the highest fractional polarization
measured being $60\%$. The maser EVPA distribution is
somewhat wider and not bimodal, and has
$\langle^{29}\chi\rangle=-47\pm22^\circ$. No significant polarization
was measured for the SO emission.

\subsubsection{W~Hya}

The P.A. range for W~Hya was too small to perform independent fits to
both EVPA and polarization fraction. We thus performed a
number of fits with fixed $P_l$ or $\chi$. However, these did not
allow for strong conclusions beyond that the mean fractional linear polarization is consistent with those found for the masers fo VX~Sgr.


\subsection{Maser Distribution}

\begin{figure}
\epsscale{1.0}
\plotone{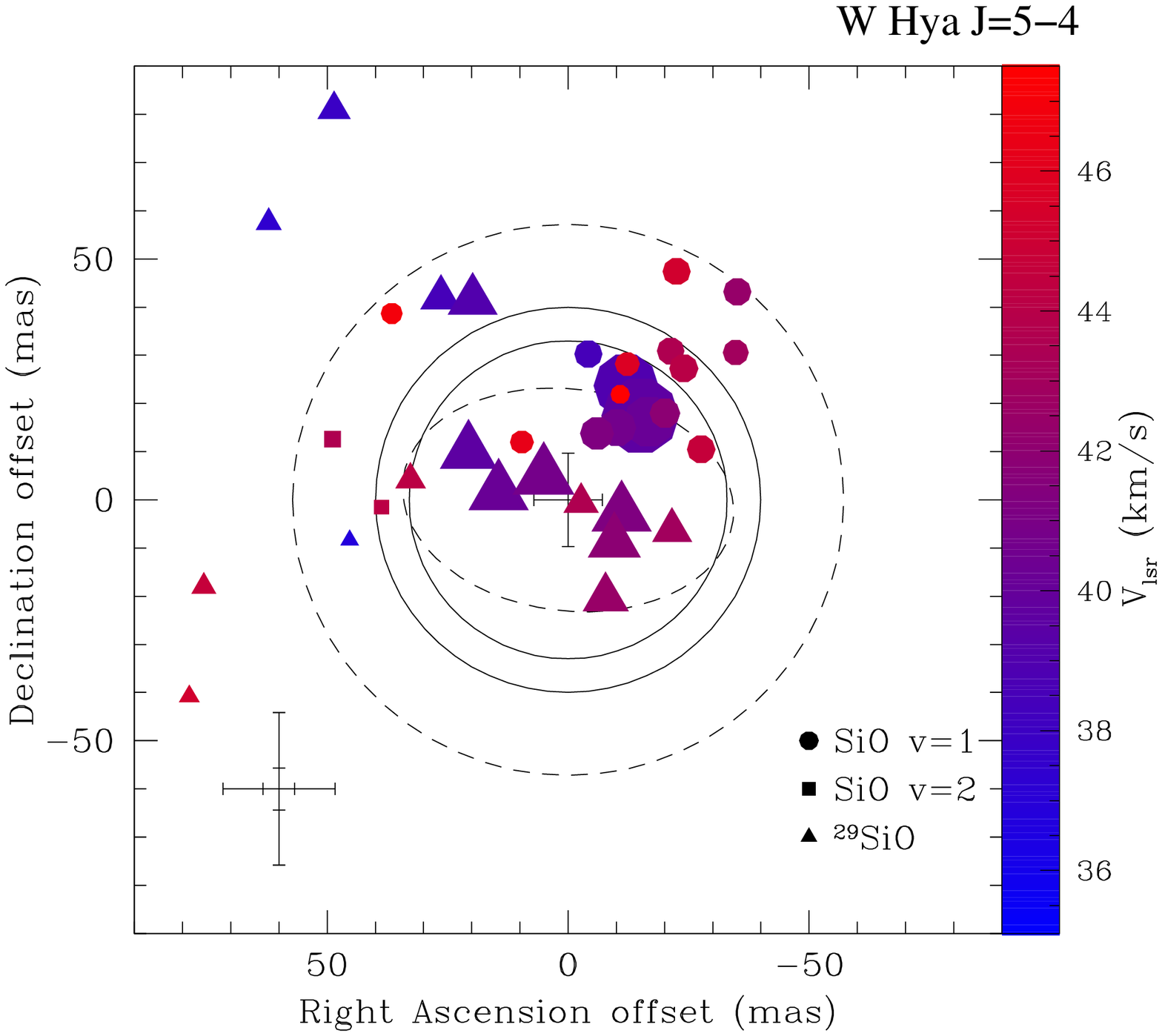}
\caption{Positions of the $J=5-4$ transitions of $^{29}$SiO $v=0$ and $^{28}$SiO $v=1,2$ around W~Hya. Positions of the masers are with respect to the peak of the dust emission ($\alpha_{\rm J2000}=13^h49^m01^s.953\pm0.015$ and $\delta_{\rm J2000}=-28^\circ22'06".208\pm0.200$) for spots with a SNR$>20$. The dust position is given in the center with positional uncertainties. The symbols are scaled linearly according to brightness, ranging from $2.2$--$11.1$~\jyb. The minimum and maximum error bars of the plotted SiO maser spots are shown in the bottom right corner. The outer, dashed, circle indicates the inner dust shell as determined from models by \citet{Wishnow10}. The solid lines indicate the radius of the 43~GHz SiO maser observations at two epochs \citep{Cotton04, Cotton08}. The dashed ellipse is the elongated radio-photosphere as determined by \citet{Reid07}.} \label{whyasio}
\end{figure}

The maser distributions of VX~Sgr and W~Hya are shown in
Figs.~\ref{vxsio} and \ref{whyasio}. The formal position errors on
the masers are large and depend on the signal-to-noise ratio (SNR).
We therefore only present those masers detected with a high SNR, whose
positions are determined with better than $\sim10$~mas accuracy. The
distributions of the $^{28}$SiO and $^{29}$SiO masers around VX~Sgr
are presented individually
with respect to the peak of the dust emission, which was determined with an accuracy of $\sim14$~mas (corresponding to $\sim24~AU$).


  Some evidence for asymmetry in the dust envelope of VX~Sgr has been
  seen using infrared interferometry \citep{Monnier04}, and also the
  dust emission peak of the supergiant VY~CMa appears offset from the
  SiO maser emission \citep{Muller07}. However, if the dust peak of VX~Sgr
  denote the stellar position, the $^{28}$SiO maser distribution is
  quite asymmetric. The $^{29}$SiO maser spots however, do appear to
  occur around the dust peak. For W~Hya no clear offset between the
dust and the masers was seen, and the positions for all three maser
transitions are presented together with respect to the peak of the
dust emission. Still, because of possible asymmetry in the dust
emission this is not necessarily the stellar position.

\subsection{SO Emission}

\begin{figure}
\epsscale{1.0}
\plotone{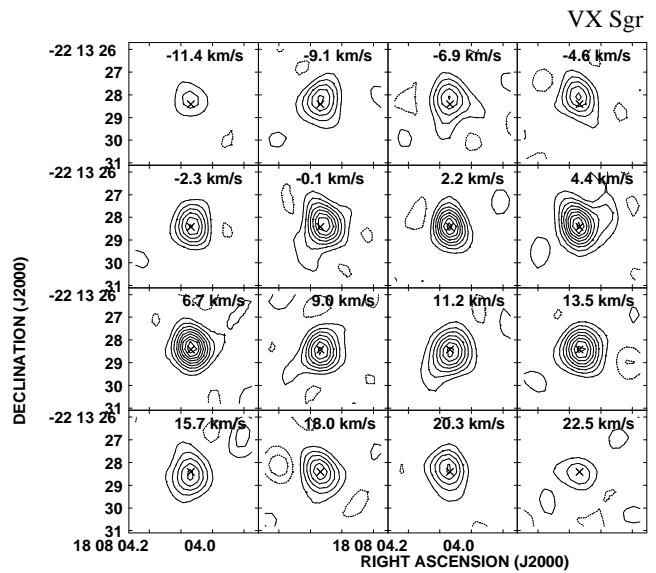}
\caption{Contour plots of the SO $5_5-4_4$ emission around VX~Sgr. The contours are drawn at 10\% levels of the peak brightness ($795$~mJy~beam$^{-1}$). The cross denotes the peak position of the dust emission.} \label{vxso}
\end{figure}

Emission from the SO $5_5-4_4$ transition at 215.2~GHz was detected
for both VX~Sgr and W~Hya and is shown in
  Fig.~\ref{spectra}. Although more typically observed at higher and
lower frequency transitions, sulfur bearing species are frequently
found in the envelopes of O-rich stars \citep[e.g.][]{Omont93}. Sulfur
is very reactive with OH and will form SO. The SO emission of both
VX~Sgr and W~Hya covers only a small velocity range around the stellar
velocity, a range comparable to that of the SiO masers. In both stars
the emission peaks near the stellar velocities with a flux density of
$\sim1$~Jy. For W~Hya, we find that, as indicated in
Fig.~\ref{whyaso}, the peak of the integrated red-shifted emission
between $40-47$~\kms~ is offset from that of the blue-shifted emission
between $33-47$~\kms. The separation between the emission peaks is
$0.29\pm0.04$~arcseconds, which at the distance of W~Hya corresponds
to $\sim28$~AU on the sky. The offset is almost exactly in the
North-South direction, with a position angle of $3^\circ\pm10$. The SO
emission of VX~Sgr does not show a significant velocity separation and
peaks at the same position as the dust as shown in Fig.~\ref{vxso}.

\subsection{Dust Continuum}

\begin{figure}
\epsscale{1.0}
\plotone{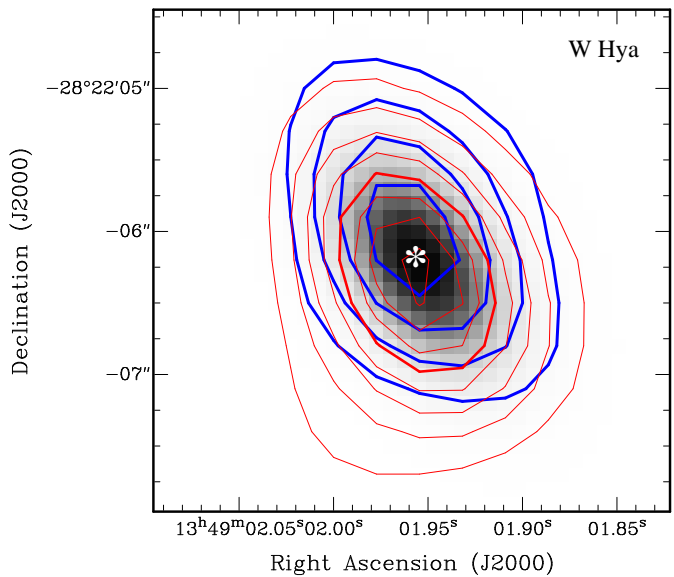}
\caption{Contours of the integrated red-shifted ($V_{\rm lsr}>40$~\kms; thin red) and blue-shifted ($V_{\rm lsr}<40$~\kms; thick blue) SO $5_5-4_4$ emission around W~Hya over-plotted on the dust emission (greyscale). Contours are drawn at intervals of $0.1$~\jyb, starting at $0.05$~\jyb~($4\sigma$). The star denotes the peak position of the dust emission.} \label{whyaso}
\end{figure}

The $\lambda=1.4$~mm continuum emission toward both W Hya
(Fig.\ref{whyaso}) and VX~Sgr is unresolved.  The integrated flux
density of the W~Hya emission is $S_{\nu}=213\pm3$~mJy, a value
comparable with previous measurements of $270\pm15$~mJy (SMA $1.1$~mm;
\citealt{Muller08}), $280\pm17$~mJy (SIMBA $1.2$~mm;
\citealt{Dehaes07}), and $280\pm30$~mJy (JCMT $1.1$~mm;
\citealt{Veen95}), taking into account a lower stellar contribution to
the flux at longer wavelength.  In the case of VX Sgr, the total flux
density $S_{\nu}=95\pm2$~mJy. To our knowledge no previous
  millimeter dust measurements of VX~Sgr are available for
  comparison.

\section{Discussion}

\subsection{The Location of the SiO Maser Transitions}

The maser distributions shown in Figs.~\ref{vxsio} and \ref{whyasio}
do not show the clear circular morphology often seen in the 43~and
86~GHz SiO masers \citep[e.g.][]{Cotton04, Soriaruiz04}. However, with
the angular resolution provided by the SMA we cannot reach the sub-mas
positional accuracy obtained with very long baseline interferometry
for the lower frequency transitions. The $^{28}$SiO maser spots
  of VX~Sgr are offset from the peak of the dust emission by $\sim
  28\pm14$~mas towards the North-West. However, the $^{29}$SiO masers
  do show a hint of a circular distribution with an inner radius of
  $\sim18$~mas, slightly more than the $\sim12$~mas observed for the
  43~GHz masers \citep{Chen06}. The situation for W~Hya is less
clear, although for this source 43~GHz masers also display only an
incomplete ring morphology \citep[e.g.][]{Cotton08}. Our observations
do let us conclude that specifically the strongest of the $^{28}$SiO
and $^{29}$SiO masers avoid each other, something that was also found
for the 43~GHz $^{28}$SiO and $^{29}$SiO masers of IRC+10011 by
\citet{Soriaruiz05}.

\subsection{Linear Polarization}

Masers can become linearly polarized under the influence of a magnetic
field \citep{Goldreich73}. Here we measured the linear polarization of
the $J=5-4$ SiO masers of VX~Sgr to be up to $\sim80\%$. This is
consistent with the linear polarization fractions measured for the
$v=1, J=5-4$ $^{28}$SiO masers of VY~CMa in S04. For the $J=1-0$ SiO
maser transitions, such polarization fractions can be reached due to
interaction with a magnetic field of a few Gauss
\citep[e.g.][]{Western84}. However, for the higher $J$ transitions, fractional
linear polarization arising solely from interaction of a saturated
maser with a magnetic field is expected to be much lower. Already for
the $J=2-1$ SiO maser transition, similar polarization fractions would
need magnetic field strengths of $>10$~G \citep{Western84} and for
$J=5-4$ transitions this will be even higher. However, in the presence
of anisotropic pumping, the linear polarization fraction can actually
increase with the angular momentum of the involved state
\citep{Nedoluha94}.

The high linear polarization is therefore largely due to
anisotropic pumping of the maser. We now need to assess if the
EVPA can still be used to derive the magnetic field
morphology. As described in for example \citet{Nedoluha90} this
depends on the ratios between the maser decay rate $\Gamma$, the maser
stimulated emission rate $R$ and the Zeeman frequency shift
$g\Omega$. Assuming a typical magnetic field of $3.5$~G
\citep{Herpin06}, the Zeeman frequency shift of the $J=5-4$ SiO masers
is $\sim800$~s$^{-1}$. The maser decay rate
$\Gamma\sim5$~s$^{-1}$. The rate for stimulated emission of the
$J=5-4$ SiO masers is given by
$R\approx4\times10^{-6}~T_b\Delta\Omega$ (S04), where $T_b$ is the
maser brightness temperature and $\Delta\Omega$ its beaming
angle. Assuming a maser spot size of $\sim0.4$~AU similar to that of
the VX~Sgr 43~GHz SiO masers \citep{Greenhill95}, the brightest masers
in our observations have $T_b\sim2\times10^9$~K and the weakest ones
have a $T_b$ almost two orders of magnitude less. Taking
$\Delta\Omega\sim10^{-2}$~sr \citep[e.g.][]{Reid88}, this yields
$R\sim80$. Therefore $g\Omega>R, \Gamma$. For these values, the results
from \citet{Nedoluha90} indicate that the linear polarization still
traces the magnetic field direction, either parallel or perpendicular,
even in the presence of anisotropic pumping. Specifically, the
polarization vectors are parallel to the magnetic field if the angle
between the field and the line-of-sight $\theta<\theta_{\rm
  crit}\approx55^\circ$ and perpendicular if $\theta>\theta_{\rm
  crit}$, where $\theta_{\rm crit}$ is the ``van Vleck angle'' \citep{Goldreich73}.

\subsection{The Shape of the Magnetic Field around VX~Sgr}

Observations of 1612~MHz OH maser polarization suggest that, in the
OH maser region at $\sim1400$~AU from VX~Sgr, the magnetic field has a
dipole shape with a position angle $\Theta=210^\circ\pm20$ and an
inclination angle $i$ between $20-30^\circ$ from the plane of the sky
\citep{Szymczak97, Szymczak01}. A similar dipole configuration, with
$\Theta=220^\circ\pm10$ and $i=40^\circ\pm5$ was found from \water
polarization observations between $\sim100-300$~AU from the star
\citep{Vlemmings05}.

We have now been able to probe the magnetic field structure at only
$\sim20-100$~AU from the star using SiO maser polarization observations. As seen
in Fig.\ref{vxsio}, the masers indicate a large scale structure in the
field with a preferred axis at $\sim30^\circ$ or $\sim-45^\circ$,
which, within the quoted uncertainties, is consistent with a
$90^\circ$ change of EVPA direction. Such a $90^\circ$
flip has previously also been observed in the 43~GHz SiO masers, and
is attributed to a change of the angle between the magnetic field and
the line-of-sight $\theta$ through $\theta_{\rm crit}$
\citep{Kemball97}. This is consistent with a magnetic field with an
inclination axis close to $\theta_{\rm crit}$, as the curvature of the
magnetic field near the star would, close to the plane of the sky
where most SiO masers are tangentially amplified, produce masers with
$\theta$ both somewhat larger and smaller than $\theta_{\rm
  crit}$. Our SiO maser observations therefore suggest that the magnetic
field close to the star has a morphology with an inclination axis
$i\sim90-\theta_{\rm crit}=35^\circ$ and a position angle
$\Theta\sim135^\circ\pm20$ or $\Theta\sim210^\circ\pm20$. These
  values are consistent with the likely dipole shaped field further
  out in the envelope, although a full 3-dimensional
  reconstruction of the field in the SiO maser region is complicated
  by the unknown location of the masers along the line of
  sight. Still, the SiO masers offer further strong evidence for a
  large-scale, most probably dipole-shaped, magnetic field around
  VX~Sgr and, considering the other maser observations, a position
  angle of $\sim210^\circ$ is most likely. Combining information from
all three maser species at $20-1400$~AU distance from VX~Sgr, we
conclude that VX~Sgr likely has a dipole magnetic field with
$\Theta=217^\circ\pm7$ and $i=37^\circ\pm9$.

Having confirmed the existence of a similar large scale magnetic field
from the SiO maser region out to the OH maser regions, we can
extrapolate the fields measured on the OH and \water masers back to
the SiO maser region at $\sim50$~AU \citep{Vlemmings05}. As, for a
dipole magnetic field, $B\propto R^{-3}$ and as the OH and \water maser
distances from the star are only roughly known, the uncertainty on this
extrapolation is large. Still, we find the field in the SiO maser
region to be in the range of $B\sim25-100$~G. While larger then the
typical field found by \citet{Herpin06}, this is comparable to the
$B=87$~G field measured in the SiO region of VX~Sgr by
\citet{Barvainis87}.

\subsection{The SO Envelope of W~Hya}

The detection of spatially offset red- and blue-shifted SO emission
around W~Hya reveals an asymmetry in the CSE. Previous high angular
resolution SO observations indicate that the SO emission can be
enhanced in collimated outflows and/or equatorial structures
\citep[e.g.][]{Trung09, Contreras00}. This implies that W~Hya harbors
either a slow bipolar outflow or a rotating and possibly expanding
equatorial disk. The position angle of the SO structure, measured
between the peaks of the red- and blue-shifted SO emission, is
$3^\circ\pm10$, nearly perpendicular to the strongly elliptical
radio-photophere measured by \citet{Reid07} to have a position angle
of $83^\circ\pm18$ and a size of $69\times46$~mas. Additionally, the
red- and blue-shifted separation in SO is similar, both in scale,
velocity and in position angle, to that observed in the OH maser
region \citep{Szymczak98}. From this it was inferred that the OH
masers originated in a weak bipolar outflow or a tilted circumstellar
disk.  Our observations of the SO emission thus further confirm this
hypothesis, although we also cannot discriminate between an outflow, a
disk or other more complex kinematic structure. This will have
to wait for higher angular resolution observations with e.g. ALMA.

\subsection{Circumstellar Dust}

\subsubsection{W~Hya}

From the observed, unresolved, dust continuum of W~Hya, we can
determine whether stellar black body or thermal dust emission
dominates the millimeter continuum. We estimate the stellar black body
contribution using the following parameters for W Hya: distance,
$D=98$~pc \citep{Vlemmings03}; stellar temperature, $T_{\rm
  eff}=2500$~K \citep{Justtanont05}; and from this stellar radius,
$R_{\rm star}=3.4\times10^{13}$~cm. Noting that $T_{\rm eff}$ and
$R_{\rm star}$ vary significantly during the stellar pulsation cycle
and have uncertainties of $\sim25-50\%$ \citep[e.g.][]{Dehaes07}, so
that this calculation is quite approximate, we estimate $S_{\rm
  star,1.4mm}=158$~mJy. Thus, as was also found by \citet{Muller08} at
$1.1$~mm, a significant component of the millimetre continuum is due
to the star itself rather than dust. Using the Rayleigh-Jeans
approximation to calculate dust mass, $M_{d}$, assuming optically thin
emission, following \citet{Muller07}

\begin{equation} 
\label{e:dustmass}
M_d = \frac{2 c^2 D^2 a_d \rho_d (S_{\rm tot}-S_{\rm star})}{3 Q_{\nu} k T_d \nu^2}
\end{equation}

\noindent where grain size $a_d=0.2\mu$m, grain mass density
$\rho_d=3.5$~g cm$^{-3}$ and emissivity $Q_{\nu}=5.65\times
10^{-4}(\nu/274.6~{\rm GHz})$, typical assumed values for oxygen-rich stars.
Adopting a beam-averaged dust temperature $T_{d}=700$~K, we obtain a dust mass
$M_{d}=4.9\times10^{-7}$~M$_{\odot}$ within a radius of about 70~AU. 

The millimeter dust observations do not probe averaged mass loss
history over the entire circumstellar envelope, but instead can give a
good indication of recent mass loss rate. For a W~Hya expansion
velocity of $V_{\rm exp}=7$~\kms \citep{Dehaes07}, and a beam of
$\sim1.5$~arcseconds, we are probing material that has been ejected only over
the past $\le50$ years. We derive a gas+dust mass loss rate of $\ge
9.8\times10^{-7}$ M$_{\odot}$yr$^{-1}$, assuming a gas to dust ratio
of 100.  This is comparable to the distance-adjusted
rate of $1.7\times10^{-6}$ M$_{\odot}$yr$^{-1}$ from ISO observations
of \citet{Zubko00}, and only slightly higher than
the distance adjusted $4\times10^{-7}$~M$_{\odot}$yr$^{-1}$ value
determined from ODIN observations by \citet{Justtanont05}.


\subsubsection{VX~Sgr}

In order to estimate the central star black body contribution to the
continuum of VX~Sgr, we adopt $D=1.57$~kpc \citep{Chen07}, $R_{\rm
  star}=1.0\times10^{14}$~cm and $T_{\rm eff}=3200$~K
\citep{Monnier04}. At $\lambda=1.4$~mm, we estimate that the stellar
component is only $S_{\rm star,1.4mm}= 6.7$~mJy. Using
Eqn.~\ref{e:dustmass}, the mass of the dust enclosed in a region of
radius $\sim1300$ AU is therefore
$M_{d}=2.1\times10^{-4}$~M$_{\odot}$. Here we have also assumed a
beam-averaged $T_{d}=700$~K, based on results of \citet{Danchi94}.
For the expansion velocity, \citet{Murakawa03} find velocities of 10
and 20~\kms at the inner and outer edges of the water maser zone at
100 and 300~AU respectively.  Adopting an average $V_{\rm
  exp}=15$~\kms, the observed region corresponds to an age of about
400 years. We estimate a total gas+dust mass of $2.1\times10^{-2}$
M$_{\odot}$ in this region, and therefore a recent mass-loss rate of
$5.3\times 10^{-5}$~M$_{\odot}$yr$^{-1}$. This is in reasonable
agreement with other mass loss measurements e.g. $3.2\times10^{-5}$
M$_{\odot}$yr$^{-1}$ \citep{Netzer89}.

\section{Conclusions}

We have determined the magnetic field morphology at a few stellar
radii from VX~Sgr using SiO masers, and find that it is consistent with the
interpretation in terms of a dipole magnetic field from OH and \water maser
observations. This is the first evidence that circumstellar magnetic field
morphology is conserved from close to the star to the outer edge of
the CSE. As the magnetic field strengths measured in the OH and \water
maser regions imply the magnetic field is dynamically important, a
further extrapolation of this dipole field to the star ($\propto
R^{-3}$) indicates that also in the SiO maser region the magnetic
field dominates the kinetic and thermal energies, with an estimated
field strength of $B\sim25-100$~G.

Our images of the SiO masers around VX~Sgr and W~Hya show that the two
$v=1,2$ rotational $^{28}$SiO transitions as well as the $^{29}$SiO
$v=0$ transition avoid each other. While a ring structure is seen in
the $^{29}$SiO line, the masers do not seem to be confined to a
narrow region.

Finally, our SO $5_5-4_4$ observations reveal the possible
presence of a slow bipolar outflow or a rotating, disk-like structure
around W~Hya.  This structure is nearly perpendicular to the
elliptically extended emission previously detected from the
radio-photophere.

High angular resolution observations of submillimeter SiO masers and
their polarization will become relatively straightforward with the
Atacama Large Millimeter/submillimeter Array (ALMA). 
Extremely interesting science targets in their own right, the strength and high
polarization fraction of the SiO masers also make them good ALMA polarization calibrators for a variety of frequency bands.

\begin{acknowledgements}
This research was supported by the Deutsche Forschungsgemeinshaft (DFG) through the Emmy Noether Research grant VL~61/3-1. 
\end{acknowledgements}


\begin{thebibliography}{}
\expandafter\ifx\csname natexlab\endcsname\relax\def\natexlab#1{#1}\fi

\bibitem[Barvainis et al.(1987)]{Barvainis87} Barvainis, R., McIntosh, G., \& Predmore, C.~R.\ 1987, \nat, 329, 613 
\bibitem[Cernicharo et al.(1993)]{Cernicharo93} Cernicharo, J., Bujarrabal, V., \& Santaren, J.~L.\ 1993, \apjl, 407, L33 
\bibitem[Chapman \& Cohen(1986)]{Chapman86} Chapman, J.~M., \& Cohen, R.~J.\ 1986, \mnras, 220, 513 
\bibitem[Chen et al.(2006)]{Chen06} Chen, X., Shen, Z.-Q., Imai, H., \& Kamohara, R.\ 2006, \apj, 640, 982 
\bibitem[Chen et al.(2007)]{Chen07} Chen, X., Shen, Z.-Q., \& Xu, Y.\ 2007, \cjaa, 7, 531
\bibitem[Chen \& Shen(2008)]{Chen08} Chen, X., \& Shen, Z.-Q.\ 2008, \apj, 681, 1574 
\bibitem[Cotton et al.(2004)]{Cotton04} Cotton, W.~D., et al.\ 2004, \aap, 414, 275 
\bibitem[Cotton et al.(2008)]{Cotton08} Cotton, W.~D., Perrin, G., \& Lopez, B.\ 2008, \aap, 477, 853
\bibitem[Danchi et al.(1994)]{Danchi94} Danchi, W.~C., Bester, M., Degiacomi, C.~G., Greenhill, L.~J., \& Townes, C.~H.\ 1994, \aj, 107, 1469 
\bibitem[Dehaes et al.(2007)]{Dehaes07} Dehaes, S., Groenewegen, M.~A.~T., Decin, L., Hony, S., Raskin, G., \& Blommaert, J.~A.~D.~L.\ 2007, \mnras, 377, 931 
\bibitem[Dinh-V.-Trung et al.(2009)]{Trung09} Dinh-V.-Trung, Muller, S., Lim, J., Kwok, S., \& Muthu, C.\ 2009, \apj, 697, 409
\bibitem[Efron \& Tibshirani(1991)]{Efron91} Efron, B., \& Tibshirani, R.\ 1991, Science, 253, 390
\bibitem[Goldreich et al.(1973)]{Goldreich73} Goldreich, P., Keeley, D.~A., \& Kwan, J.~Y.\ 1973, \apj, 179, 111  
\bibitem[Gomez Balboa \& Lepine(1986)]{Gomez86} Gomez Balboa, A.~M., \& Lepine, J.~R.~D.\ 1986, \aap, 159, 166 
\bibitem[Gray et al.(1999)]{Gray99} Gray, M.~D., Humphreys, E.~M.~L., \& Yates, J.~A.\ 1999, \mnras, 304, 906 
\bibitem[Greenhill et al.(1995)]{Greenhill95} Greenhill, L.~J., Colomer, F., Moran, J.~M., Backer, D.~C., Danchi, W.~C., \& Bester, M.\ 1995, \apj, 449, 365
\bibitem[Herpin \& Baudry(2000)]{Herpin00} Herpin, F., \& Baudry, A.\ 2000, \aap, 359, 1117 
\bibitem[Herpin et al.(2006)]{Herpin06} Herpin, F., Baudry, A., Thum, C., Morris, D., \& Wiesemeyer, H.\ 2006, \aap, 450, 667
\bibitem[Ho et al.(2004)]{Ho04} Ho, P.~T.~P., Moran, J.~M., \& Lo, K.~Y.\ 2004, \apjl, 616, L1
\bibitem[Humphreys et al.(1997)]{Humphreys97} Humphreys, E.~M.~L., Gray, M.~D., Yates, J.~A., \& Field, D.\ 1997, \mnras, 287, 663 
\bibitem[Humphreys(1999)]{Humphreys99} Humphreys, E.~M.~L.\ 1999, Ph.D.~Thesis, Bristol University, UK
\bibitem[Jewell et al.(1987)]{Jewell87} Jewell, P.~R., Dickinson, D.~F., Snyder, L.~E., \& Clemens, D.~P.\ 1987, \apj, 323, 749 
\bibitem[Justtanont et al.(2005)]{Justtanont05} Justtanont, K., et al.\ 2005, \aap, 439, 627 
\bibitem[Kemball \& Diamond(1997)]{Kemball97} Kemball, A.~J., \& Diamond, P.~J.\ 1997, \apjl, 481, L111 
\bibitem[Kemball \& Martinsek(2005)]{Kemball05} Kemball, A., \& Martinsek, A.\ 2005, \aj, 129, 1760
\bibitem[Kemball(2007)]{Kemball07} Kemball, A.~J.\ 2007, IAU Symposium, 242, 236
\bibitem[Kemball et al.(2010)]{Kemball10} Kemball, A., Martinsek, A., Mitra, M., \& Chiang, H.-F.\ 2010, \aj, 139, 252
\bibitem[Lockett \& Elitzur(1992)]{Lockett92} Lockett, P., \& Elitzur, M.\ 1992, \apj, 399, 704 
\bibitem[Monnier et al.(2004)]{Monnier04} Monnier, J.~D., et al.\ 2004, \apj, 605, 436 
\bibitem[Muller et al.(2007)]{Muller07} Muller, S., Dinh-V-Trung, Lim, J., Hirano, N., Muthu, C., \& Kwok, S.\ 2007, \apj, 656, 1109 
\bibitem[Muller et al.(2008)]{Muller08} Muller, S., Dinh-V-Trung, He, J.-H., \& Lim, J.\ 2008, \apjl, 684, L33
\bibitem[Murakawa et al.(2003)]{Murakawa03} Murakawa, K., Yates, J.~A., Richards, A.~M.~S., \& Cohen, R.~J.\ 2003, \mnras, 344, 1 
\bibitem[Nedoluha \& Watson(1990)]{Nedoluha90} Nedoluha, G.~E., \& Watson, W.~D.\ 1990, \apjl, 361, L53 
\bibitem[Nedoluha \& Watson(1994)]{Nedoluha94} Nedoluha, G.~E., \& Watson, W.~D.\ 1994, \apj, 423, 394 
\bibitem[Netzer(1989)]{Netzer89} Netzer, N.\ 1989, \apj, 342, 1068 
\bibitem[Omont et al.(1993)]{Omont93} Omont, A., Lucas, R., Morris, M., \& Guilloteau, S.\ 1993, \aap, 267, 490
\bibitem[Pardo et al.(1998)]{Pardo98} Pardo, J.~R., Cernicharo, J., Gonzalez-Alfonso, E., \& Bujarrabal, V.\ 1998, \aap, 329, 219 
\bibitem[Reid \& Menten(2007)]{Reid07} Reid, M.~J., \& Menten, K.~M.\ 2007, \apj, 671, 2068 
\bibitem[Reid \& Moran(1988)]{Reid88} Reid, M.~J., \& Moran, J.~M.\ 1988, Galactic and Extragalactic Radio Astronomy, 255 
\bibitem[S{\'a}nchez Contreras et al.(2000)]{Contreras00} S{\'a}nchez Contreras, C., Bujarrabal, V., Neri, R., \& Alcolea, J.\ 2000, \aap, 357, 651
\bibitem[Soria-Ruiz et al.(2004)]{Soriaruiz04} Soria-Ruiz, R., Alcolea, J., Colomer, F., Bujarrabal, V., Desmurs, J.-F., Marvel, K.~B., \& Diamond, P.~J.\ 2004, \aap, 426, 131
\bibitem[Shinnaga et al.(2004)]{Shinnaga04} Shinnaga, H., Moran, J.~M., Young, K.~H., \& Ho, P.~T.~P.\ 2004, \apjl, 616, L47 (S04) 
\bibitem[Soria-Ruiz et al.(2005)]{Soriaruiz05} Soria-Ruiz, R., Colomer, F., Alcolea, J., Bujarrabal, V., Desmurs, J.-F., \& Marvel, K.~B.\ 2005, \aap, 432, L39 
\bibitem[Szymczak \& Cohen(1997)]{Szymczak97} Szymczak, M., \& Cohen, R.~J.\ 1997, \mnras, 288, 945 
\bibitem[Szymczak et al.(1998)]{Szymczak98} Szymczak, M., Cohen, R.~J., \& Richards, A.~M.~S.\ 1998, \mnras, 297, 1151
\bibitem[Szymczak et al.(2001)]{Szymczak01} Szymczak, M., Cohen, R.~J., \& Richards, A.~M.~S.\ 2001, \aap, 371, 1012 
\bibitem[van der Veen et al.(1995)]{Veen95} van der Veen, W.~E.~C.~J., Omont, A., Habing, H.~J., \& Matthews, H.~E.\ 1995, \aap, 295, 445 
\bibitem[Vlemmings et al.(2003)]{Vlemmings03} Vlemmings, W.~H.~T., van Langevelde, H.~J., Diamond, P.~J., Habing, H.~J., \& Schilizzi, R.~T.\ 2003, \aap, 407, 213
\bibitem[Vlemmings et al.(2005)]{Vlemmings05} Vlemmings, W.~H.~T., van Langevelde, H.~J., \& Diamond, P.~J.\ 2005, \aap, 434, 1029
\bibitem[Vlemmings(2007)]{Vlemmings07} Vlemmings, W.~H.~T.\ 2007, IAU Symposium, 242, 37
\bibitem[Western \& Watson(1984)]{Western84} Western, L.~R., \& Watson, W.~D.\ 1984, \apj, 285, 158 
\bibitem[Wishnow et al.(2010)]{Wishnow10} Wishnow, E.~H., Townes, C.~H., Walp, B., \& Lockwood, S.\ 2010, \apjl, 712, L135
\bibitem[Zubko \& Elitzur(2000)]{Zubko00} Zubko, V., \& Elitzur, M.\ 2000, \apjl, 544, L137 
\end{thebibliography}
\end{document}